\documentclass[a4paper,11pt]{article}
\usepackage{pos}
\usepackage{multirow}
\usepackage{sidecap}
\usepackage{feynmp-auto}
\usepackage{adjustbox}
\usepackage{comment}

\newcommand{\be}{\begin{equation}}
\newcommand{\ee}{\end{equation}}
\newcommand\beq{\begin{eqnarray}}
\newcommand\eeq{\end{eqnarray}}

\title{Toward a resolution of the NN controversy}

\author*[a,i]{~Amy Nicholson,}
\author[n]{~Evan Berkowitz,}
\author[j]{~John Bulava,}
\author[i,h,p]{~Chia Cheng Chang,}
\author[k]{~M.A. Clark,}
\author[f]{~Andrew D. Hanlon,}
\author[b,i]{~Ben H\"orz,}
\author[c,i]{~Dean Howarth,}
\author[g]{~Christopher K\"orber,}
\author[l]{~Wayne Tai Lee,}
\author[i,h]{~Aaron S. Meyer,}
\author[o]{~Henry Monge-Camacho,}
\author[m]{~Colin Morningstar,}
\author[d,e,p,q]{~Enrico Rinaldi,}
\author[c,i]{~Pavlos Vranas}
\author[i,h,c]{~and Andr\'e Walker-Loud}

\affiliation[a]{Department of Physics and Astronomy, University of North Carolina,
  Chapel Hill, NC 27516-3255, USA}

\affiliation[b]{Intel Corporation,
Santa Clara, CA 95054, USA}

\affiliation[c]{Physics Division, Lawrence Livermore National Laboratory,
Livermore, CA 94550, USA}
\affiliation[d]{Physics Department, University of Michigan, Ann Arbor, MI 48109, USA}
\affiliation[e]{Theoretical Quantum Physics Laboratory, RIKEN, 2-1 Hirosawa, Wako, Saitama 351-0198, Japan}

\affiliation[f]{Physics Department, Brookhaven National Laboratory, Upton, New York 11973, USA}

\affiliation[g]{Institut f\"ur Theoretische Physik II, Ruhr-Universit\"at Bochum, D-44780 Bochum, Germany}

\affiliation[h]{Department of Physics, University of California,
Berkeley, CA 94720, USA}

\affiliation[i]{Nuclear Science Division, Lawrence Berkeley National Laboratory,
Berkeley, CA 94720, USA}

\affiliation[j]{CP3 -Origins \& Dept. of Mathematics and Computer Science, University
of Southern Denmark Campusvej 55,
5230 Odense M, Denmark}

\affiliation[k]{NVIDIA Corporation,
Santa Clara, CA 95050, USA}

\affiliation[l]{Department of Statistics, Columbia University,
New York, NY 10027, USA}

\affiliation[m]{Department of Physics, Carnegie Mellon University,
Pittsburgh, PA 15213, USA}
\affiliation[n]{Institut f\"ur Kernphysik and Institute for Advanced Simulation, Forschungszentrum J\"ulich,
54245 J\"ulich, Germany}

\affiliation[o]{Escuela de Física, Universidad de Costa Rica, San Jos\'e, San Pedro, 11501, Costa Rica}
\affiliation[p]{iTHEMS, RIKEN, 2-1 Hirosawa, Wako, Saitama 351-0198, Japan}
\affiliation[q]{Center for Quantum Computing, RIKEN, 2-1 Hirosawa, Wako, Saitama 351-0198, Japan}
\emailAdd{annichol@email.unc.edu}

\abstract{Lattice QCD calculations of two-nucleon interactions have been underway for about a decade, but still haven't reached the pion mass regime necessary for matching onto effective field theories and extrapolating to the physical point. Furthermore, results from different methods, including the use of the L\"uscher formalism with different types of operators, as well as the HALQCD potential method, do not agree even qualitatively at very heavy pion mass. We investigate the role that different operators employed in the literature may play on the extraction of spectra for use within the L\"uscher method. We first explore expectations from Effective Field Theory solved within a finite volume, for which the exact spectrum may be computed given different physical scenarios. We then present preliminary lattice QCD results for two-nucleon spectra calculated using different operators on a common lattice ensemble.}

\FullConference{%
 The 38th International Symposium on Lattice Field Theory, LATTICE2021
  26th-30th July, 2021
  Zoom/Gather@Massachusetts Institute of Technology
}

\begin{document}
\maketitle

\section{Introduction}
Lattice QCD presents a unique opportunity to understand nuclear physics from first principles, making accurate calculations of two-nucleon systems of critical importance for benchmarking computational techniques to be used for many-body systems. In particular, the construction of optimal two-nucleon operators and the faithfulness of non-perturbative methods for extracting physical multi-particle observables must be well-understood before calculations of larger systems may be trusted. Furthermore, reliable calculations of two-nucleon scattering are necessary for the extraction of unknown two-nucleon matrix elements of interest for low-energy beyond the Standard Model searches, such as neutrinoless double beta decay. 

To date, two-nucleon calculations have been performed at unphysically large pion masses by several groups. However, the results stemming from different computational methods have shown stark discrepancies, giving even qualitatively different pictures of physics at heavy pion mass (see Ref.~\cite{Drischler:2019xuo} for a more detailed discussion on this issue). On the other hand, results from different techniques should only be expected to agree if systematic effects are properly accounted for. It has been suggested that the most likely source of error in approaches which utilize spectroscopy for extracting scattering information is excited-state contamination in the determination of the energies~\cite{Iritani:2017wvu,Iritani:2018vfn}. Therefore, a controlled study of the different methods on a single ensemble, where systematics such as discretization effects are expected to be equivalent, should help to enlighten the situation. 

Here we present steps toward understanding the operator dependence of the extraction of two-nucleon spectra for use with the L\"uscher method of calculating two-nucleon scattering phase shifts. We examine, in particular, the energies extracted using a large basis of momentum-space operators in combination with variational techniques, as well as more traditional position-space to momentum-space correlation functions with the two-nucleons at the source placed at a single spacetime location, or spatially separated by some distance. These calculations have all been performed on a single CLS ensemble having $m_{\pi} \sim 714$~MeV. We furthermore explore the overlap of the various operators onto two-nucleon states within an effective field theory (EFT) framework. Here we can tune the scattering phase shift to a variety of possibilities, and determine the exact finite-volume spectrum that would result from such a phase shift. We then form correlation functions using different operators and analyze the extracted spectrum versus the exact spectrum. 

\section{Operators and excited states}

Historically, the cheapest method computationally for calculating two-nucleon correlation functions has been to create the two nucleons on the same spatial lattice point at the source, and project onto various non-interacting momentum levels at the sink. This method, pioneered by NPLQCD and used extensively to calculate two- and multi-nucleon observables, predicts that at unphysically large pion mass two-nucleon systems form bound states in both $s$-wave channels, with a binding energy much larger than that of the physical deuteron~\cite{NPLQCD:2013bqy,NPLQCD:2012mex}. More recently, however, sophisticated variational calculations by multiple groups have favored unbound systems at similarly heavy pion masses~\cite{Green:2021qol,Horz:2020zvv,Francis:2018qch,Amarasinghe:2021lqa}. The difference between these results likely does not signal a failure of the L\"uscher method which is common to all these calculations, but rather, to differences in the spectra determined in each case. While all operators used are expected to decay to the same ground state energy for sufficiently large Euclidean time, the well-known signal-to-noise problem with nucleons prohibits calculations at such large times. Thus, one must be able to correctly identify energy levels at relatively short Euclidean times, with an accurate assessment of the systematics associated with excited-state contamination. 

The lowest-lying excited states can be roughly categorized as stemming from inelastic, single-nucleon excitations, such as a nucleon-pion state, or elastic two-body scattering states arising from the finite size of the lattice volume. The energy gap between inelastic excited states thus scales roughly as $m_{\pi}$, while elastic excited states scale inversely with the spatial extent of the box and the nucleon mass. These elastic excited-state energy gaps are therefore much smaller than those for inelastic excitations, and contaminations from these states can vary extremely slowly with Euclidean time. Variational methods attempt to project out contributions from a single energy level by diagonalizing a correlation matrix created from a large basis of operators, each having different overlap onto a given state. This basis may include, for example, nucleons at different non-interacting momentum levels, or systems of two nucleons having different single-nucleon wavefunctions, or smearings. In order to project out individual elastic scattering states, one would expect to require an operator basis having different spatial locations or momentum projections included.

In order to illustrate the different overlaps of two-nucleon operators onto different \textit{elastic} excited states, one may use Effective Field Theory (EFT) as input. Chiral Perturbation Theory, for example, has been used to illustrate excited-state effects in single-nucleon observables (see, e.g., Ref.~\cite{Bar:2017}). 
In the two-nucleon sector at very low energies (below the mass of the pion), two nucleons interacting in a box obey a pionless EFT, and their scattering phase shifts encode all the information about the two-body wavefunctions for the interacting systems in a box. 
Because the nucleons are treated as point particles, inelastic excited-state contamination may not be studied within this EFT. However, for the lowest-energy excitations in the box - elastic excited states which do not probe the details of the nucleons themselves - the wavefunctions calculated within the pionless EFT should be an accurate approximation of those of the fully interacting two-nucleon systems of QCD.

\section{Excited-state spectra from EFT}

Because the exact two-nucleon phase shift at heavier-than-physical pion masses is not known, we tune the EFT to reproduce multiple physical scenarios, namely, one in which a deep bound state is the true ground state of the system, and one in which there is no bound state at all. We then produce correlation functions within this (discretized) EFT using various types of operator setups and examine the spectra that one would extract at intermediate time ranges (relative to the box length L, which sets the scale for excited-state energy splittings). 

The EFT in use is based off the works of ~\cite{Chen:2003vy,Endres:2012cw}, and is a discretized version of pionless EFT, in which point-like non-relativistic nucleon fields, $\psi(x)$, interact via delta-function interactions,
\beq
\mathcal{L}_{\mathrm{eff}} = \psi^{\dagger} \left(i\partial_t + \frac{\nabla^2}{2M}\right) \psi + g_0 \left(\psi^{\dagger}\psi \right)^2 \ .
\eeq
The discretized version leads to the following two-particle transfer matrix in momentum-space,
\beq
\langle pq| \mathcal{T} | p' q' \rangle = \frac{\delta_{pp'}\delta_{qq'}+\frac{g_0}{V} \delta_{p+q,p'+q'}}{\sqrt{\xi(p)\xi(q)\xi(q')\xi(p')}} \ , \qquad \xi(p) \equiv \frac{\Delta(q)}{M}  \ ,
\eeq
where $\Delta(q)$ is a chosen lattice dispersion relation for the nucleon fields, $M$ is the nucleon mass, and $V$ is the spacetime volume. This transfer matrix is projected onto the cubic irreps for $s$-wave scattering.

The coupling, $g_0$, may be tuned to reproduce infinite-volume scattering phase shifts having a variety of low-energy properties. Correlation functions in time, $t$, are then created by taking $t$ products of this transfer matrix, then projecting at source and sink times to various operators. Position space operators may be created via Fourier Transform of the transfer matrix at a given time slice. In this section we will explore the extracted spectra for attractive interactions leading to two different physical scenarios: no bound state, and a deeply-bound state (binding energy much larger than the typical energy gaps between scattering states).

\subsection{\label{sec:nobound}No physical bound state}

\begin{figure}
\begin{center}
\includegraphics[width=0.42\linewidth]{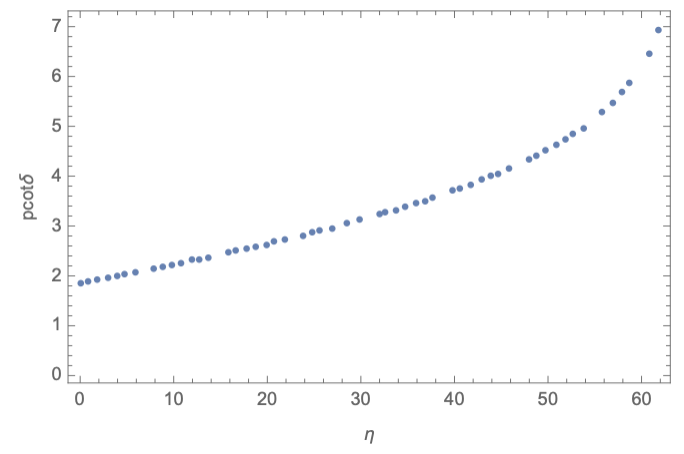}
\includegraphics[width=0.45\linewidth]{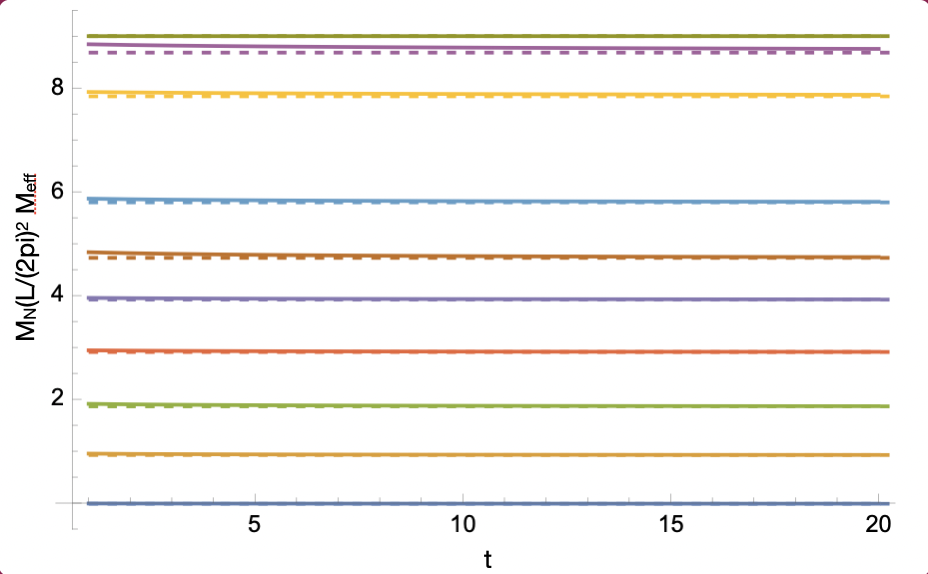}
\end{center}
\caption{\label{fig:1}Left: Scattering phase shift in lattice units versus momentum, $\eta = \left(\frac{p L}{2\pi}\right)^2$ for the first system discussed in Sec.~\ref{sec:nobound} (having no physical bound state). Right: Effective masses for the 10 lowest-lying two-nucleon states in an $L=12$ box as solved from a GEVP with a basis of 10 momentum-space two-nucleon operators. The exact spectrum is shown as a set of dashed lines.}
\end{figure}

In Figure~\ref{fig:1} (left), we show the exact scattering phase shift, $p \cot \delta$, versus the scattering momentum, $\eta \equiv \left(\frac{p L}{2\pi}\right)^2$, evaluated within our EFT for a chosen coupling, $g_0$. On the right is shown the exact spectrum (horizontal dashed lines), as calculated within a finite volume. Also shown at right are the effective masses of the solutions to a generalized eigenvalue problem (GEVP), solved using a correlation matrix composed of the operators corresponding to the 10 lowest-lying non-interacting momentum states of the two-particle system in the box. One sees that even for very short times relative to the inverse excited-state energy gap (note the rescaling of the vertical axis) and using a small basis, this variational method correctly reproduces the exact spectrum, up to the next-to-highest state. 

Other methods previously used for two-nucleon calculations involve the use of position-space operators at the source, with momentum-projected operators at the sink. The utility of these methods is the reduced computational time associated with the generation of relatively few quark propagators. For example, the local hexaquark operator used by the NPLQCD collaboration for the majority of their multi-nucleon calculations requires the generation of only one light quark propagator for each type of single-nucleon operator used. However, these hexaquark operators have roughly equal overlap onto all possible non-interacting momentum scattering states in the box, requiring the sink operators to effectively separate out individual states in the interacting system. Because the overlaps of source and sink are not the same, and in fact, nearly orthogonal, cancellation between contributions from different excited states can occur, leading to the potential for what has been dubbed a ``fake plateau": a relatively stable result over an intermediate time range that does not correspond to the true ground state of the system, but rather a local maximum or minimum in the effective mass (see, e.g., Ref.~\cite{Iritani:2017wvu}). 

\begin{figure}
\begin{center}
\includegraphics[width=0.45\linewidth]{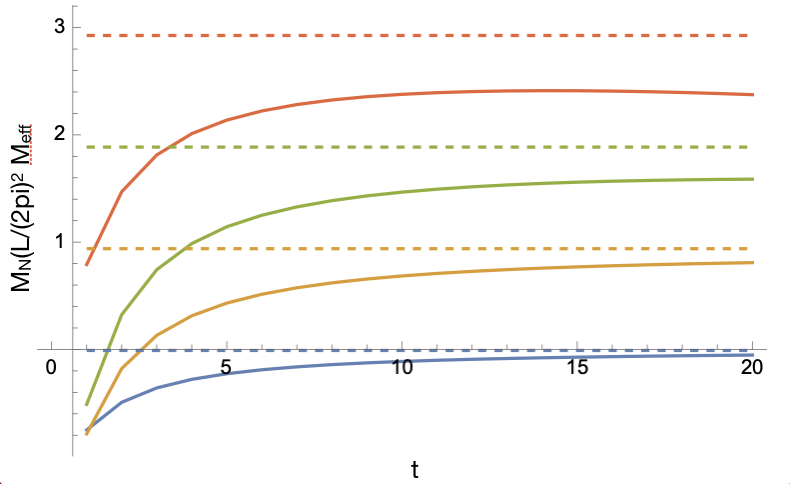}
\includegraphics[width=0.45\linewidth]{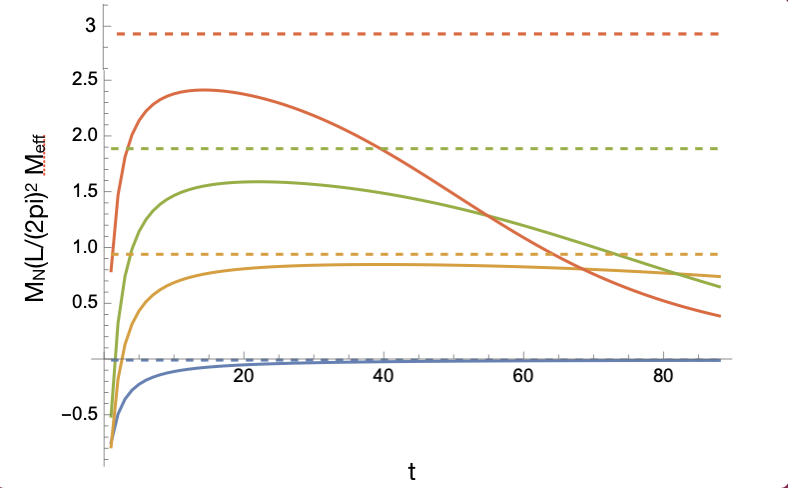}
\includegraphics[width=0.45\linewidth]{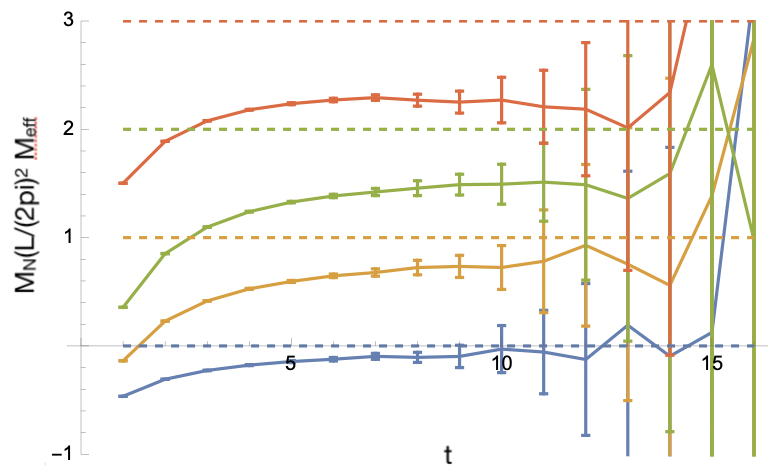}
\includegraphics[width=0.45\linewidth]{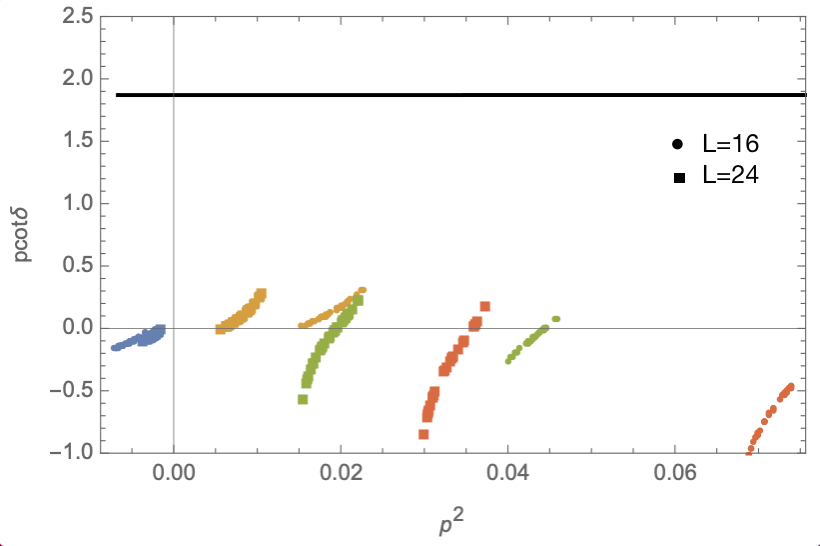}
\end{center}
\caption{\label{fig:2}Effective masses for the two-nucleon system in an for the scattering phase shift shown in Fig.~\ref{fig:1} (no physical bound state) using local hexaquark two-nucleon operators at the source and momentum-space operators at the sink up to intermediate times (upper left) and late times (upper right). The exact spectrum is shown as a set of dashed lines. On the bottom panels are shown the same effective masses with uncorrelated, exponentially growing Gaussian noise (left), as well as fits to this data processed through the L\"uscher formalism to give an effective scattering phase shift versus momentum (lower right). In this phase shift plot, different colors correspond to the different states extracted, with blue corresponding to the ground state, yellow the first excited state, etc. The process is repeated for multiple volumes (corresponding to different markers on the plot) to obtain the full data set.}
\end{figure}

This is indeed what is seen in the EFT. In Figure~\ref{fig:2}, we show the effective mass as a function of time for the same system as shown for the variational method, but calculated using local operators at the source and momentum space operators at the source (upper left). One sees a relative leveling out of the effective masses for intermediate times, however, the values are systematically found at values lower than the exact energies (dashed lines). This perceived leveling off can clearly be seen to originate from a set of local maxima when viewed at much larger times (upper right), where all correlation functions approach the ground state. When uncorrelated exponentially growing gaussian noise is added to the system (lower left), any subtle time dependence of these false states is obscured; this effect may be further enhanced by the correlated time fluctuations seen in real lattice QCD calculations. The ground state effective mass is seen to approach the true ground state energy very slowly from below. In a full lattice QCD calculation, the single-nucleon correlation function approaches the ground state from above. Thus, the combination of inelastic excited states from the single nucleon at early times and slowly varying elastic excited states at late times would very likely conspire at intermediate times to give a perceived ground state energy at a lower value than the true ground state. 

Finally, one is not able to deduce that there is anything wrong with the extracted spectrum upon converting these incorrect energies into phase shifts using the L\"uscher method. On the lower right panel are shown the phase shift data extracted from fits to the correlation functions in the lower left, as well as those extracted using a larger volume. Both the finite-volume spectrum, and the extracted phase shift at low energies predict a bound state in the system, and there are no ``smoking gun" outlier states. This may be due to the momentum projection performed at the sink. This projection allows one to reasonably separate states into the appropriate number required to give a physical phase shift, while the local operator at the source bends them each systematically down toward the ground state. 

\begin{figure}
\begin{center}
\includegraphics[width=0.45\linewidth]{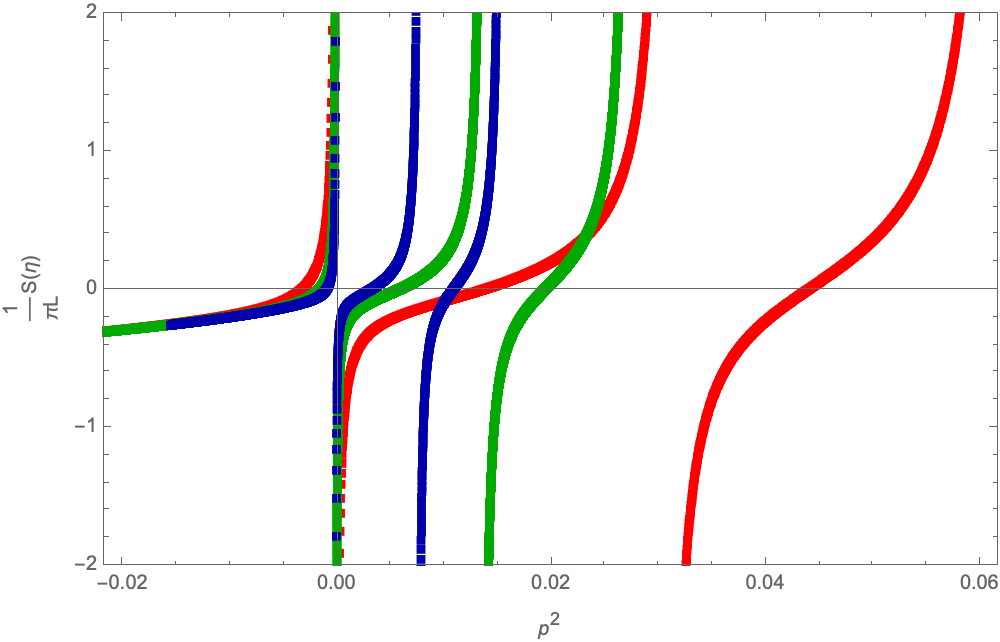}
\end{center}
\caption{\label{fig:7}Predicted phase shift versus finite-volume spectrum from the L\"uscher formalism for three different volumes, $L=16,24,32$ (red, green, blue).}
\end{figure}

Note, in particular, that the extracted ground state shows very little volume dependence (despite having no true bound state in the system), while the higher scattering states show a strong volume dependence. This may be explained by inspecting Fig.~\ref{fig:7}, where the finite-volume spectrum is shown on the horizontal axis. If the system is weakly interacting, then the true energy spectrum for a given box size will be found near the poles of the corresponding $S$-function (vertical axis). Even for an attractive, purely scattering system (with no bound states), the ground states from different volumes will show very little volume dependence, likely imperceptible given the statistics with which these states can be determined, while those states above threshold have dramatic volume dependence (compare how closely the L\"uscher curves for different volumes lie to each other below threshold compared to those above threshold). If the action of the hexaquark source is to pull the energies extracted in each volume away from the true eigenstates by roughly the same percentage, one would still expect to see little to no volume dependence below threshold and clear power-law dependence above threshold. Thus, one should not read too much into the claim that little to no volume-dependence of the ground state is consistent with a bound state; it would, in fact, require a volume-\textit{dependent} artifact of the choice of operators to change this somewhat universal behavior.

\begin{figure}
\begin{center}
\includegraphics[width=0.42\linewidth]{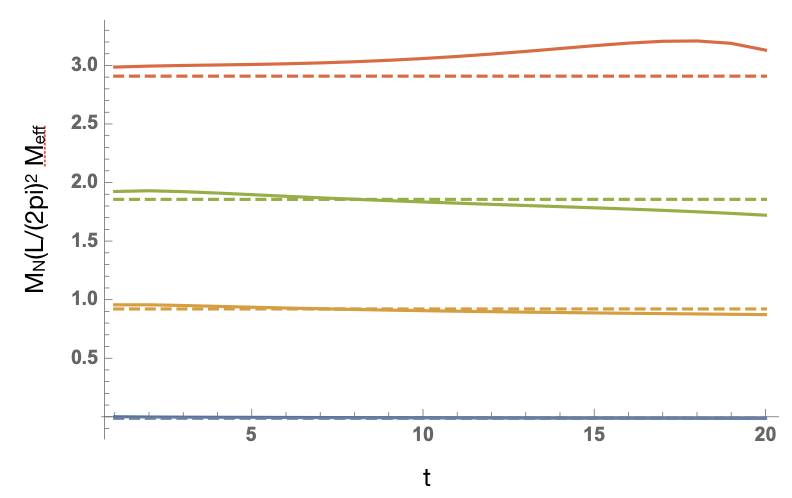}
\includegraphics[width=0.45\linewidth]{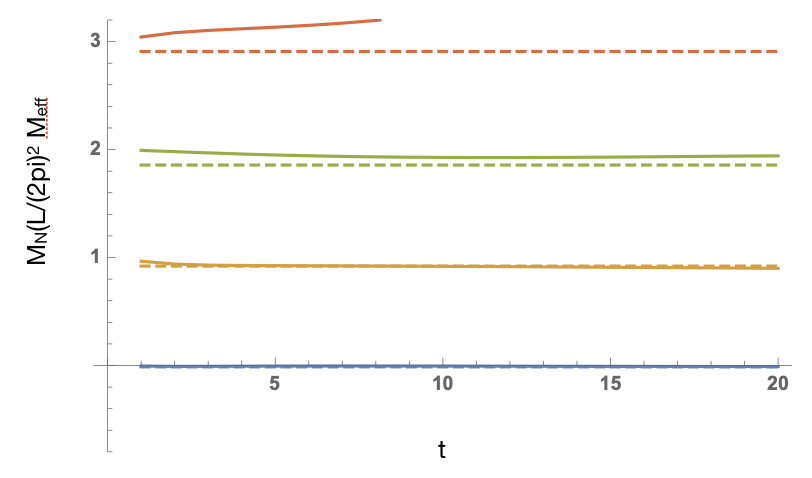}
\end{center}
\caption{\label{fig:3}Effective masses for the two-nucleon system for the scattering phase shift shown in Fig.~\ref{fig:1} (no physical bound state) using two-nucleon operators spatially displaced by $\Delta = 1$ (left) and $\Delta = 5$ (right) at the source and momentum space operators at the sink. The exact spectrum is shown as a set of dashed lines.}
\end{figure}

Our final operator study is for the displaced position-space nucleon operators at the source and momentum projected operators at the sink introduced by the CalLat collaboration in Ref.~\cite{Berkowitz:2015eaa}. The idea behind the spatial displacement was that low energy two-nucleon scattering states likely have more support from operators which are spread out in space. The results using this method for the first physical scenario are shown in Fig.~\ref{fig:3} using two-nucleon operators, $N(x)N(x+\Delta)$, displaced by $\Delta=1$ (left) and $\Delta=5$ (right) lattice sites at the source. One finds that in general the ground state is correctly reproduced, while the higher-lying states can show some dependence on the spatial displacement; this dependence may in principle be used as a measure for the remaining systematic. 

\subsection{\label{sec:bound}Deeply-bound state}

\begin{figure}
\begin{center}
\includegraphics[width=0.4\linewidth]{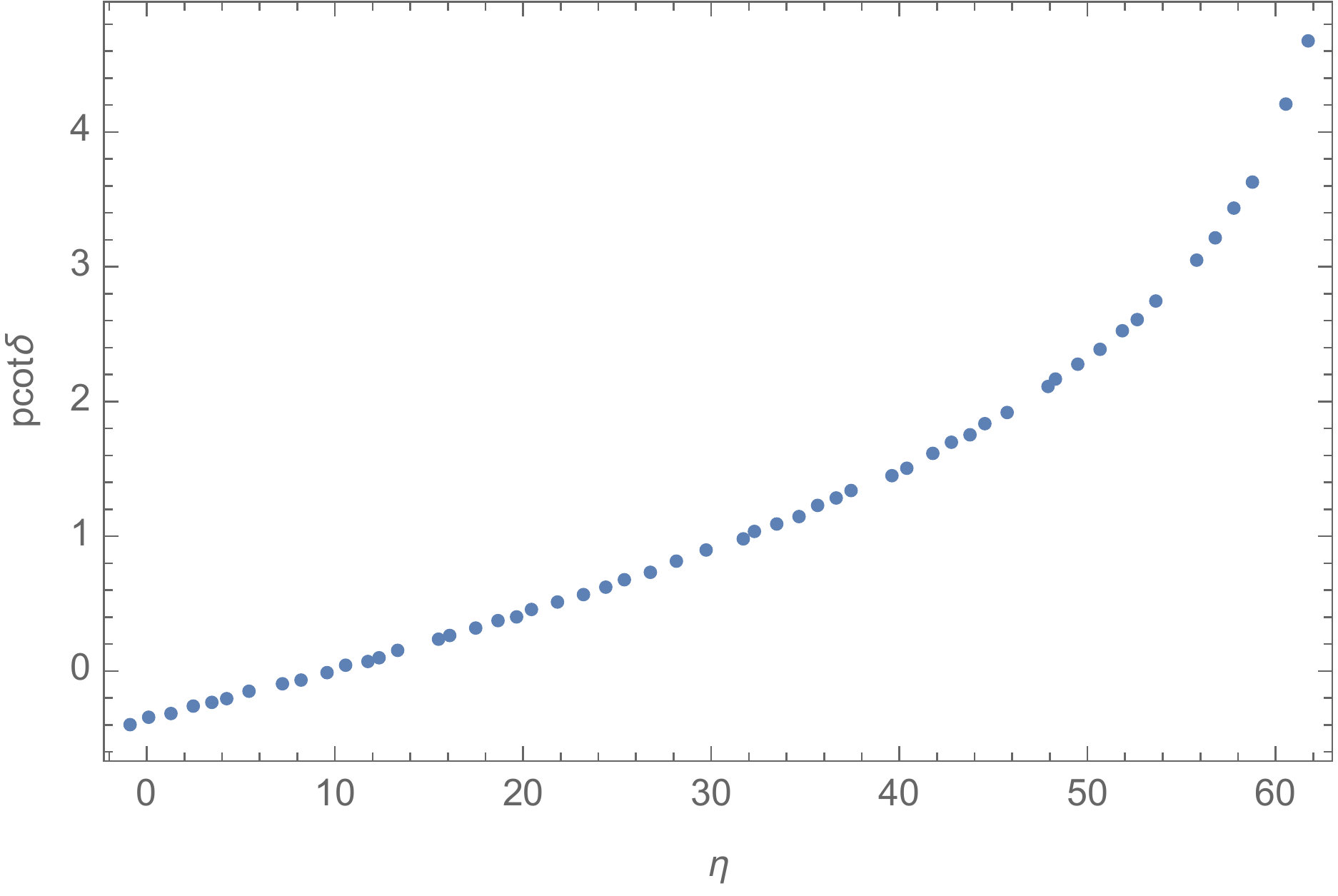}
\includegraphics[width=0.45\linewidth]{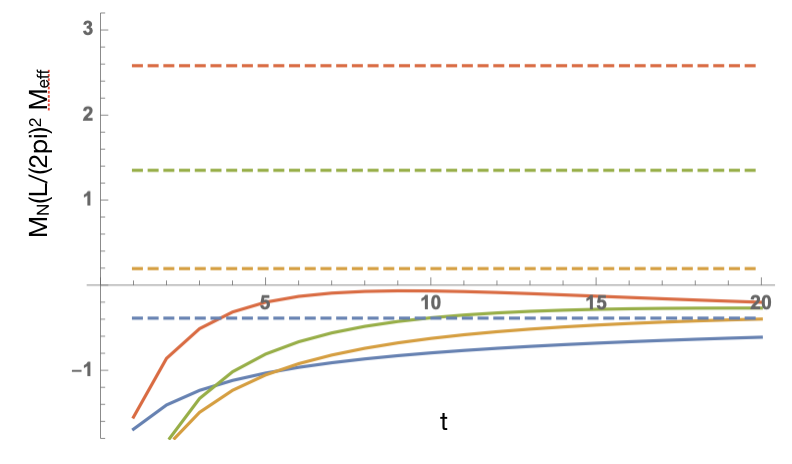}
\includegraphics[width=0.45\linewidth]{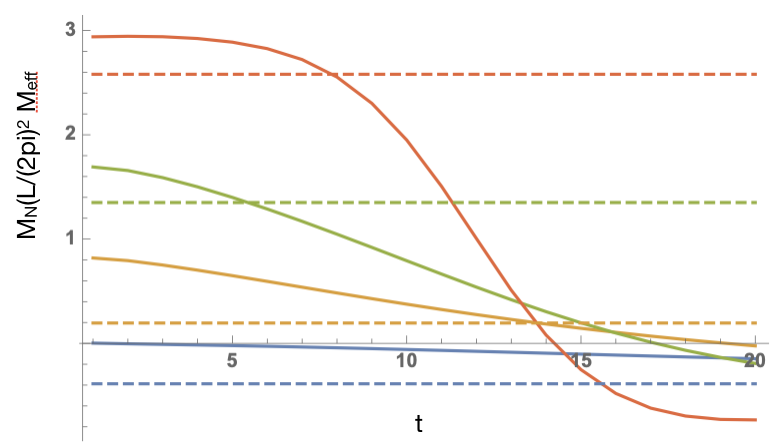}
\includegraphics[width=0.45\linewidth]{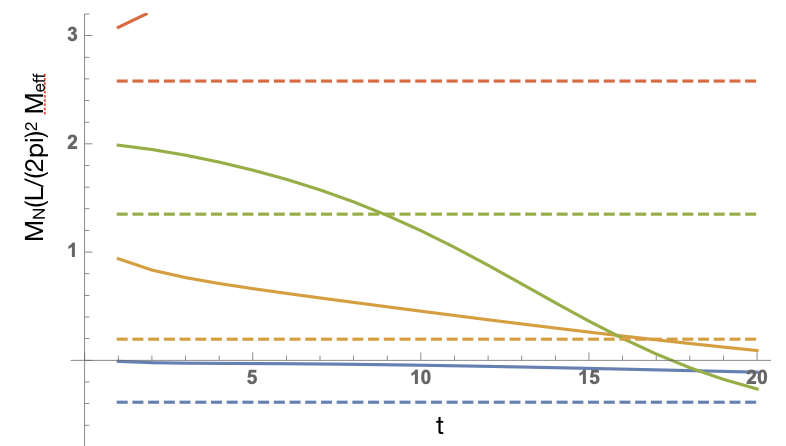}
\end{center}
\caption{\label{fig:4}Upper left: Scattering phase shift in lattice units versus momentum, $\eta = \left(\frac{p L}{2\pi}\right)^2$ for the second system discussed in Sec.~\ref{sec:bound}, having a deeply-bound ground state. Upper right: Effective masses for the two-nucleon system in an $L=12$ box using local hexaquark two-nucleon operators at the source and momentum-space operators at the sink. Lower panels: Effective masses for the two-nucleon system in an $L=12$ box using two-nucleon operators spatially displaced by $\Delta = 1$ (left) and $\Delta = 5$ (right) at the source and momentum space operators at the sink. The exact spectrum in these panels is shown as a set of dashed lines.}
\end{figure}

For this next discussion, we tune our EFT to a system containing a deep bound state (Fig.~\ref{fig:4}, upper left), and explore the results given by various operator methods. Here, we begin with the results using local hexaquark operators (Fig.~\ref{fig:4}, upper right). We see that regardless of the momentum projected onto at the sink, all correlation functions collapse  to the same (ground) state. This reinforces the intuition that local operators approximate the wavefunction of a bound state, but have relatively poor overlap onto scattering states. In this case, the local operators overlap so strongly onto the ground-state wavefunction that the momentum operator sinks are not sufficient to distinguish any of the excited scattering states in the box. Interestingly, the zero relative momentum sink operator approaches the ground state very slowly, and in practice (with noisy data) would likely lead to a false plateau at a lower energy than the true ground state, while the higher shells approach it more quickly. The CalLat displaced operators (Fig.~\ref{fig:4}, bottom), on the other hand, do project onto a correct number of distinct states at early times, however, these originate near values systematically \textit{higher} than the exact spectrum, and within intermediate times most of these have been pulled down toward the ground state.

\begin{figure}
\begin{center}
\includegraphics[width=0.45\linewidth]{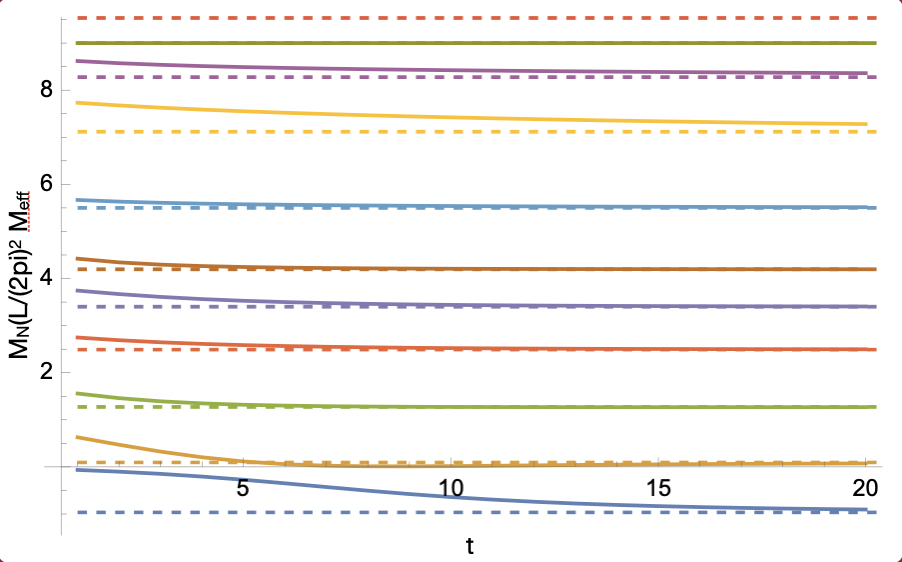}
\includegraphics[width=0.45\linewidth]{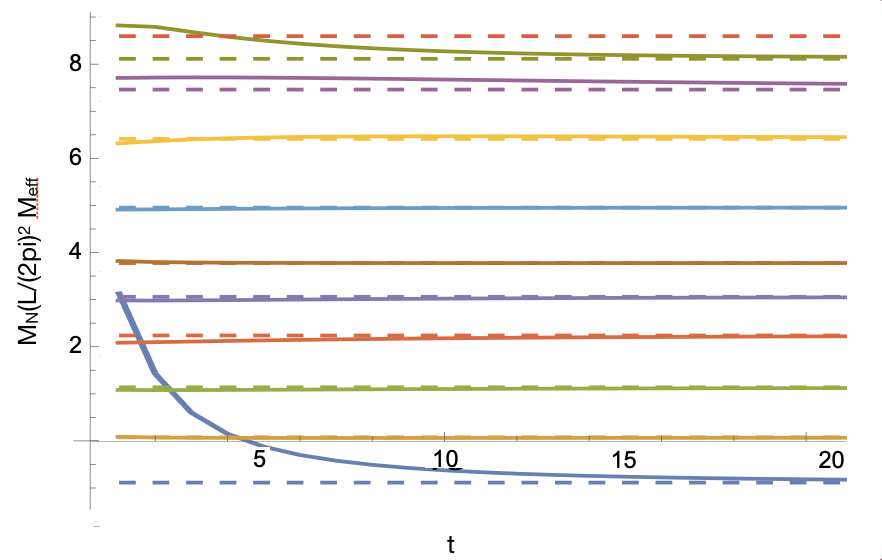}
\includegraphics[width=0.45\linewidth]{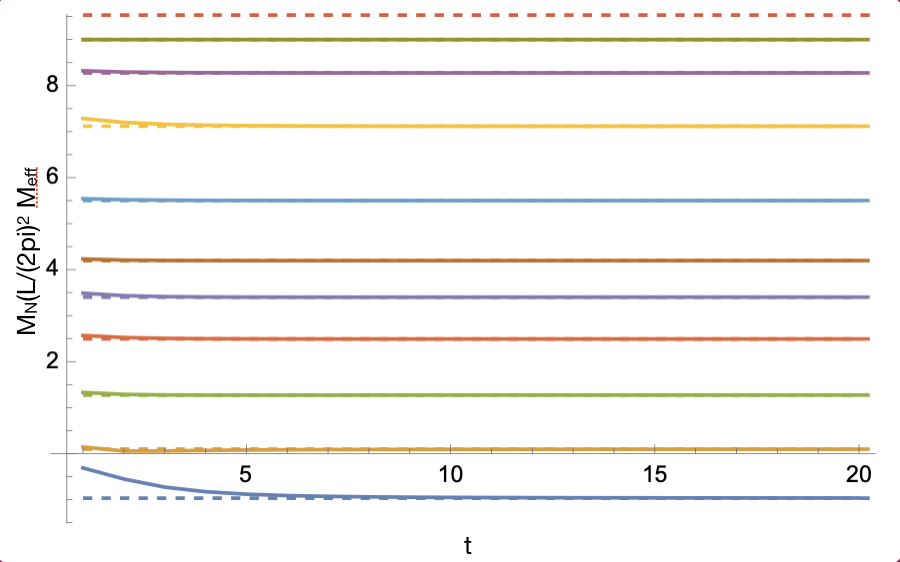}
\end{center}
\caption{\label{fig:5}Effective masses for the two-nucleon system for the scattering phase shift shown in Fig.~\ref{fig:4} (deeply-bound state) as solved from a GEVP with a basis of 10 momentum-space two-nucleon operators (upper left), 10 momentum-space operators plus a local hexaquark operator (upper right), and 30 momentum-space operators (bottom). The exact spectrum is shown as a set of dashed lines. }
\end{figure}

Results using the variational method with ten momentum-space operators is shown in Fig.~\ref{fig:5} (upper left). It is clear that this method gives the correct number of clearly separated states which reach the correct values within intermediate times. The ground state, however, is approached more slowly compared to the case when there was no bound state in the system. It should be noted, however, that because the higher-lying states are correctly described, a phase shift analysis of this data would correctly identify the bound state. Furthermore, one simply has to increase the basis of operators in order to reach the ground state (as well as the higher-lying states) more quickly (Fig.~\ref{fig:5}, bottom). One might think that adding a local hexaquark operator to the variational basis might allow one to reduce the overall basis size needed to project out the ground state at early times. However, this EFT study finds that adding the hexaquark operator does not affect either the spectrum, or the time range necessary to extract the various states (Fig.~\ref{fig:5}, upper right).  

\section{Further systematics within LQCD}

Recently, it was shown by the Mainz collaboration that large discretization effects, on the order of $\sim$30~MeV, may be present in the $H$-dibaryon system at an $SU(3)$-symmetric point with $m_{\pi}\sim 420$~MeV~\cite{Green:2021qol}. Thus, it is crucial that different methods for extracting the phase shifts be compared on a single lattice ensemble, such that other systematics are held relatively fixed. In particular, the finite-volume spectrum at a fixed lattice spacing, volume, and pion mass must not depend on the operator methods used, unless one or more of these methods contains un-quantified systematics. 

\begin{figure}
\begin{center}
\includegraphics[width=0.45\linewidth]{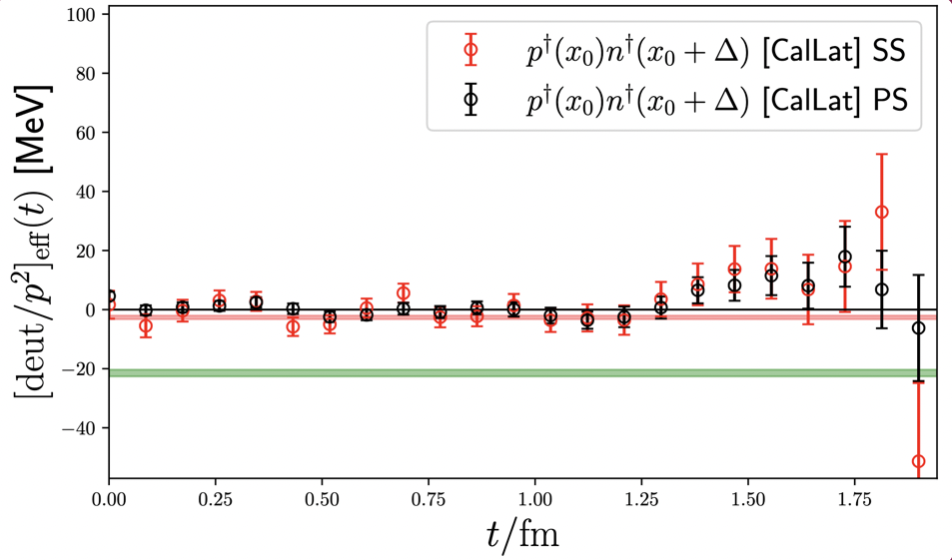}
\includegraphics[width=0.45\linewidth]{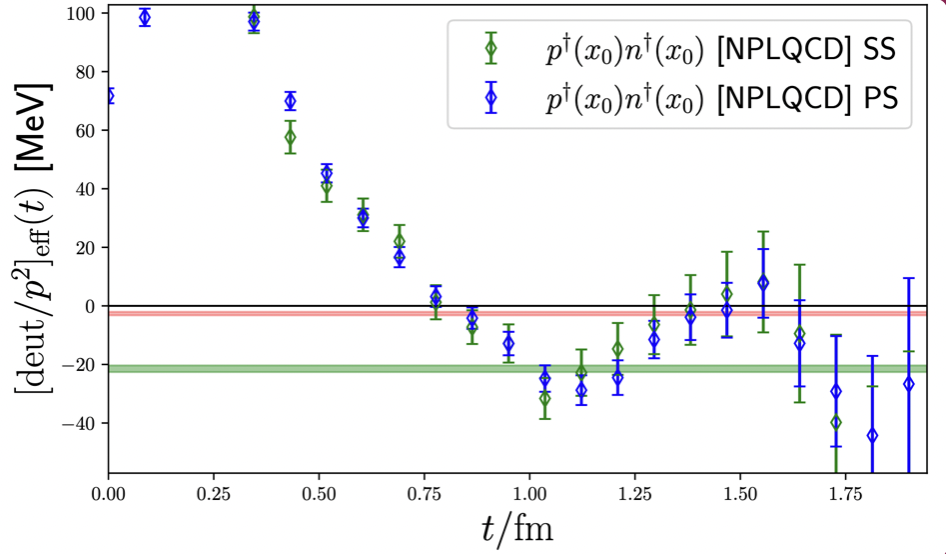}
\end{center}
\caption{\label{fig:6}Lattice QCD effective mass plots for the ground state of two-nucleons in the deuteron channel calculated using spatially displaced two-nucleon operators at the source (left) and local hexaquark operators at the source (right). In both cases, momentum-space operators are used at the sink, and two different single-nucleon wavefunctions at the source, smeared and point, are shown in different colors. Also shown are the results on the same ensemble reported in Ref.~\cite{Horz:2020zvv} calculated using a variational basis (red band), as well as the results from a different ensemble but at similar pion mass and local hexaquark operators, Ref.~\cite{NPLQCD:2013bqy} (green band).}
\end{figure}

In a first step toward this goal, we have generated correlation functions using various methods in the literature on a single CLS ensemble, with $m_{\pi}\sim 714$~MeV, $a \sim 0.086$~fm, and $L\sim 4.1$~fm. The results for the spectrum using the variational method sLapH on this ensemble were reported in~\cite{Horz:2020zvv}, where no evidence for a bound state was found in either two-nucleon channel. In Figure~\ref{fig:6}, we show preliminary results which compare the effective masses for the ground state in the deuteron channel calculated using the CalLat displaced operators (left) and the NPLQCD hexaquark operators (right) to the extracted ground state energy using sLapH (red band). We see that the correlation functions obtained using displaced operators decay almost immediately to the same ground state energy, much as was seen in Fig.~\ref{fig:3}, while the hexaquark operator decays more slowly, and appears to undershoot the true ground state energy. For comparison, we also plot the extracted ground state energy using hexaquark operators on a different ensemble, but at a similar pion mass, from Ref.~\cite{NPLQCD:2013bqy,NPLQCD:2012mex}. These observations, together with the expectations from EFT, seem to indicate that there is no true bound state in this system.

\section{Conclusions}

Excited-state systematics are a leading cause for the discrepancies in the literature for two-nucleon scattering phase shifts evaluated using different methods. Comparing the results of different ratio correlators in Fig.~\ref{fig:6}, we can conclude that the ratios give nearly perfect cancellation of single-nucleon, inelastic excited-state contamination (left panel), while the differences between the two panels stem from elastic, two-nucleon excited states. Thus, the excited-state contamination relevant for two-nucleon systems is indeed elastic, and the EFT method presented here is a qualitatively accurate description of  excited states arising due to different operator choices in lattice QCD results. By tuning the EFT to reproduce different possible physical scenarios, it is found that the variational method employing momentum-space only operators faithfully reproduces the finite-volume spectrum in each case at early and/or intermediate times using a relatively small operator basis. 

On the other hand, operator choices employing position space two-nucleon fields at the source with momentum-projected sinks likely suffer from problems. Spatially displaced two-nucleon operators work quite well for the case where there is no deeply-bound state in the system, but give incorrect spectra in the latter case. Local hexaquark operators only couple to the ground state for the case with a deeply-bound state, and lead to systematically incorrect spectra at intermediate times for essentially all cases. This implies that older lattice QCD results which utilize hexaquark operators (NPLQCD, Yamazaki et al, and hexaquark results from CalLat) are likely incorrect. Interestingly, lattice QCD results for two-nucleon systems using local hexaquark operators show well-separated spectra due to the projection onto momentum states at the sink; from this study we see that qualitatively this is the behavior expected from a physical system which contains no physical bound state, the opposite of the conclusion that has been drawn from the spectrum extracted using this method. 

Other systematics which may be at play in lattice QCD calculations include discretization effects and contributions from inelastic, single-nucleon excited states. In order to isolate only those systematics due to excited-state contamination, results from various operators should be compared on a single lattice ensemble. Preliminary results for such a study seem to agree with the EFT findings for a system with no physical bound state. It is also confirmed that spatially displaced operators agree well with the variational results at very early times.

\acknowledgments
This work was supported in part by the NVIDIA Corporation (MAC), the Alexander von Humboldt Foundation through a Feodor Lynen Research Fellowship (CK), the RIKEN Special Postdoctoral Researcher Program (ER), the Nuclear Physics Double Beta Decay Topical Collaboration (HMC, AN, AWL), the U.S. Department of Energy, Office of Science, Office of Nuclear Physics under Award Numbers DE-AC02-05CH11231 (CCC, CK, BH, AWL), DEAC52-07NA27344 (DH, PV), DE-FG02-93ER-40762 (EB), DE-SC00046548 (ASM); the DOE Early Career Award Program (AWL), and the National Science Foundation CAREER Award Program (AN). ADH is supported by the U.S. Department of Energy, Office of Science,
Office of Nuclear Physics through the Contract No. DE-SC0012704 and
within the framework of Scientific Discovery through Advance Computing
(SciDAC) award "Computing the Properties of Matter with Leadership
Computing Resources." 

\bibliographystyle{physrev} 
\bibliography{NN} 

\end{document}